\documentstyle[pre,aps,eqsecnum,graphicx]{revtex}

\title{Simulations of a mortality plateau in the sexual Penna model for biological ageing}

\author{V.~Schw\"ammle$^{1,2}$ and S.~Moss de Oliveira$^{1}$}
\address{$^1$Instituto de F\'isica, Universidade Federal Fluminense, Av. Litor\^anea, s/n - Boa Viagem, 24210-340, Niter\'oi, RJ, Brasil.}
\address{$^2$Institute for Computer Physics, University of Stuttgart,Pfaffenwaldring 27, D-70569 Stuttgart, Germany.}

\begin{document}

\maketitle

\begin{abstract}
The Penna
model is a strategy to simulate the genetic dynamics of age-structured
populations, in which the individuals genomes are represented by bit-strings. It provides 
a simple metaphor for the evolutionary process in terms of the mutation accumulation theory.
In its original version, an individual dies due to inherited diseases when 
its current number of accumulated mutations, $n$, reaches a threshold value, $T$. 
Since the number of accumulated diseases increases with age, 
the probability to die is zero for very young ages ($n < T$) and equals 1 for the old ones 
($n \ge T$).   
Here, instead of using a step function to determine the genetic 
death age, we test several other functions that may or may not slightly increase the death 
probability at young ages ($n < T$), but that decreases this probability at old ones. 
Our purpose is to study the oldest old effect, that is, a plateau in the mortality curves 
at advanced ages. Imposing certain conditions, it has been possible to obtain a clear plateau 
using the Penna model. However, a more realistic one 
appears when a modified version, that  
keeps the population size fixed without fluctuations, is used. We also find a 
relation between the birth rate, the age-structure of the population and  
the death probability. 
\end{abstract}
  
\section{Introduction}
\label{sec:intro}

The mechanism of ageing is still an important task of recent research and  
the mutation accumulation hypothesis is one of the most acceptable ones. 
The mortality can be measured by the so called mortality function: 
\begin{equation}
\mu(x) = -d\ln(S(x))/dx,
\end{equation}
where $S(x)$ is the probability to survive from birth to age $x$. Still in the 19th century 
Gompertz found that the mortality function increases exponentially with age.
Less or more pronounced decreases of this mortality exponential growth at old ages, 
also known as the oldest old effect, have been observed in humans and 
mainly, in flies~\cite{Vaupel98}.
 
The Penna 
model~\cite{Penna95,Moss99} is one of the most popular models for biological ageing, which 
has been successfully applied to 
reproduce the Gompertz law~\cite{Penna95,Moss95b},
to study the preference of sexual to asexual reproduction~\cite{Sa98} and more recently, 
to study sympatric speciation~\cite{Sa2001,Luz2003,Sousa2004}. Its asexual version 
was solved  
analytically by Coe et al~\cite{Coe2002}, who show that the replacement of a sudden death rule 
after the accumulation of $T$ deleterious mutations (step function) by a probability to  
die given by a Fermi-function, 
leads to a plateau in the mortality curve. Numerically, this plateau was only slightly 
observed before 
\cite{Thoms95,Huang2001}. 

In our simulations of the original sexual Penna model, we obtain that when the genetic 
death probability at advanced ages   
is given by a smooth function instead of the usual step function, but is   
greater than zero at very young ages, the birth rate 
has to be extremely increased to avoid population meltdown. In this case it is very difficult 
to measure the oldest old effect, since very few individuals survive until old ages.   
A small plateau has been observed by imposing a death probability equal to zero for 
very young ages ($n < T$), as in the original Penna strategy. 

Using a model where the 
population size is constant without fluctuations, we obtain a very clear plateau in 
the mortality curves, even considering non-zero values for the probability to die at 
young ages. We also obtain 
that the age distribution of the population changes dramatically according to the smoothness 
of the death probability functions at old ages. 

In Section 2 the original Penna model is briefly explained,   
the genetic death probability functions that are used in order to study the oldest 
old plateau are introduced and the corresponding results are presented. 
In Section 3 we describe the model in which 
the population size is kept constant and show the results obtained for the same death 
probability functions of the previous section. 
In Section 4 we present the conclusions.

\section{The Penna model for sexual populations}
\label{sec:orig}

In this section only a short description of the Penna model is given. A more detailed 
one can be found in~\cite{Moss99}. 
In the original version of the model two strings of 32 bits that are read in parallel 
represent the diploid genome of an individual. A deleterious mutation is defined 
by two set bits 
at the same position of both strings or by a single set bit at a dominant position.  
The dominant positions are randomly chosen at the beginning of the simulation and 
remain fixed. 
At every iteration or ``year'' one more bit 
position becomes active and the corresponding individual becomes one year older. It  
dies for genetic reasons if its current number of 
deleterious mutations reaches the threshold $T$, which corresponds to the 
following genetic death probability, $f(n)$:
\begin{equation}
\label{eq:step}
        f(n) = \Theta(n-T),
\end{equation}
where $n$ is the current number of deleterious mutations and $\Theta(x)$ is the step or 
Heavyside function.
In order to limit the population size, $P$, an additional death probability,  
$V = \frac{P(t)}{P_{max}}$, the so called Verhulst factor, is used to keep the population 
size below $P_{max}$. It is applied to each individual independently of its age or genome.  

At every iteration, any female with age equal or above the minimum reproduction age, 
$R$, randomly 
chooses a male, also with age $\ge R$, to breed and generate $b$ offspring. 
The bit-strings of the mother are crossed and recombined to produce a female gamete (single 
bit-string). One deleterious mutation is then introduced at a random position. The same process 
occurs with the father bit-strings and the union of the two gametes form the offspring genome.  
 
\subsection{Approximations to the step function:}

\begin{figure}[htb]
  \begin{center}
    \includegraphics[width=0.6\textwidth]{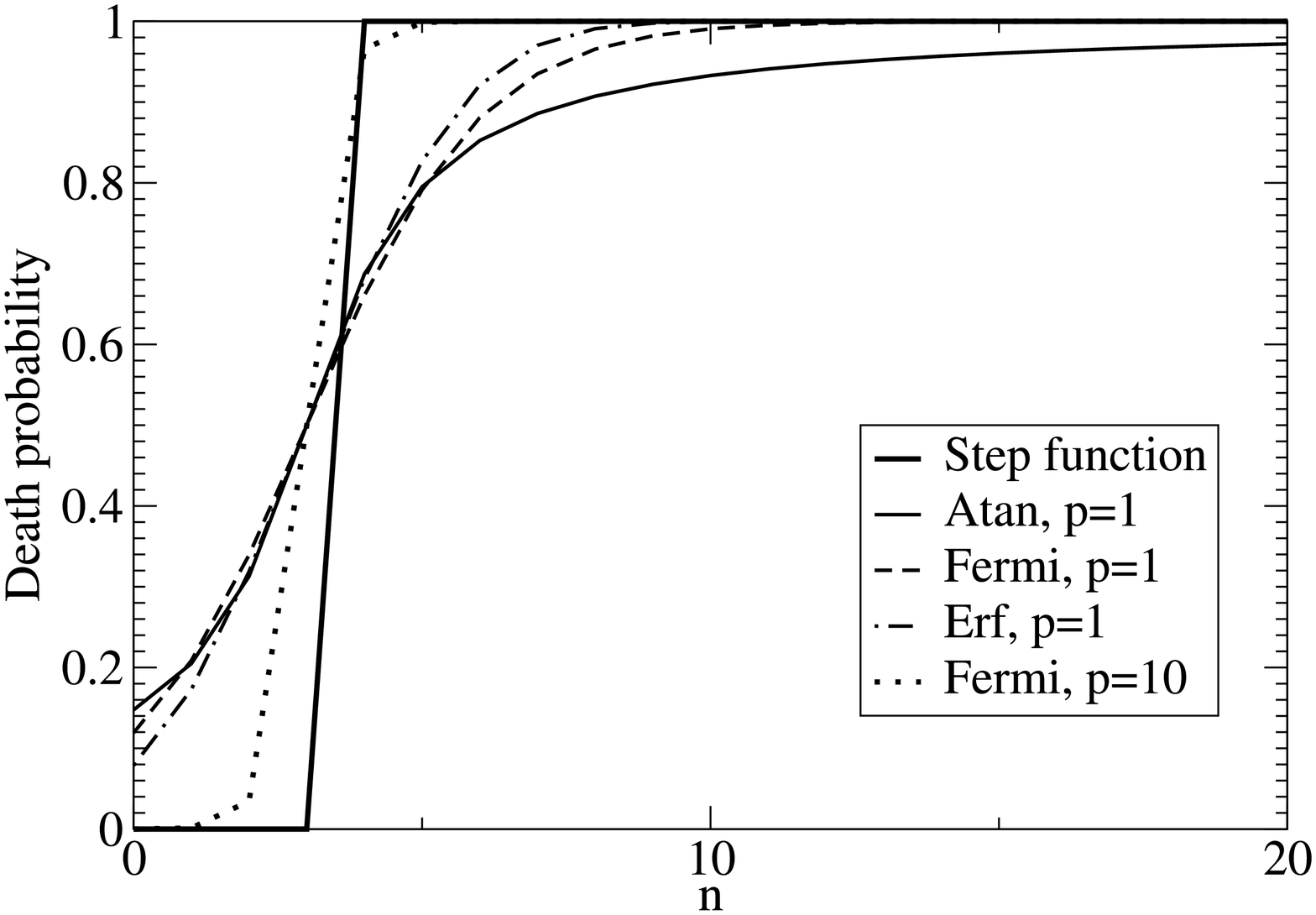}
\caption{Death probabilities according to the different approximations of the step function 
given by Equations~(\ref{eq:fermi}) to (\ref{eq:erf}), versus 
the number of accumulated diseases. The parameter $p$ controls the smoothness of the 
functions. The step function case (Equation~(\ref{eq:step})) is presented for comparison. 
Notice that for $p=1$ there is a finite probability for the very young (small $n$) to die,  
but the probability for the older to die is smaller than that given by the step function.   
For $p=10$, the behavior of the Fermi death probability becomes almost equivalent to the step 
function one.}
    \label{fig:stepf}
\end{center}
\end{figure}

We use the following approximations of the step function, 
in order to smooth the original genetic death rule of killing the individual after 
the accumulation of exactly $T$ deleterious mutations:
\begin{equation}
  {\rm Fermi \, \, function} = f_1(n) = \frac{1}{1+e^{-2p(n-T)/32}},
\label{eq:fermi}
\end{equation}
\begin{equation}
  {\rm Arc tangent \, \, function} = f_2(n) = \frac{\arctan(2p(n-T)/32)}{\pi}+\frac{1}{2},
\label{eq:atan}
\end{equation}
\begin{equation}
  {\rm Error \, \, function} = f_3(n) = \frac{1}{2} \cdot [\text{erf}(p(n-T)/32) +1],
\label{eq:erf}
\end{equation}
where $n$ is the number of active deleterious mutations and $p$ is   
a parameter that controls the smoothness of the approximations. When the 
value of $p$ increases, the death probabilities given by Equations~(\ref{eq:fermi}) to 
(\ref{eq:erf}) converge to the one given by the step function (Equation~(\ref{eq:step})). 
Figure~\ref{fig:stepf} compares the three 
approximations with $p=1$ and also the Fermi function with $p=10$, with the   
step function death probability.

\subsection{Results}
\label{sec:res}

In simulations of $N$ time steps, the mortality function, $m(a)$, is measured over 
the last $N_m$ time steps, in the following way:
\begin{equation}
m(a) = -\ln(1-\frac{\sum_{t = N_m}^{N}D_{gen}(t,a+1)}{\sum_{t = N_m}^{N}P(t,a)}),
\label{eq:label}
\end{equation}
where $D_{gen}(t,a)$ is the number of genetic deaths (not by Verhulst) at age $a$ and  
time step $t$, and $P(t,a)$ is the number of individuals with age $a$ at time step $t$. 

In all simulations that follow, the values of the parameters are: $T=3$, $R=10$, $b=1$, 
$P_{max}=200,000$, $N=100,000$, $N_m=50,000$ and the number of 
randomly chosen positions where the bits 1 are dominant is 5. 
\bigskip

The mortalities obtained using any of the death probabilities given by Equations~(\ref{eq:fermi}) to (\ref{eq:erf}) with $p \ge 10$, are equivalent to those obtained 
with the traditional step function, that is, no plateau appears.   
Smaller values of the smoothness $p$ lead to population meltdown. This can be 
avoided by increasing the birth rate $b$ to extremely high values, which produces strong 
fluctuations in the population size, or by setting the 
reproduction age to $R=1$. In fact, the latter strategy was used in \cite{Coe2002} 
to obtain the plateau. Such condition may be realistic for flies, but 
certainly not for humans. 
In fact, only simulations with very large populations and long 
simulation times, with $R=8$ showed a small plateau. 

\begin{figure}[htb]
  \begin{center}
    \includegraphics[width=0.6\textwidth]{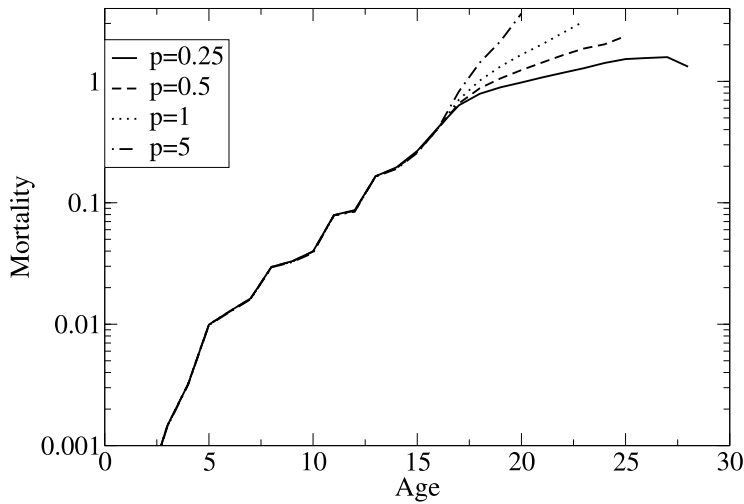}
\caption{Comparing the mortality functions for different values of $p$, using a Fermi 
death probability function. The results for small values of $p$ are similar to those 
of the analytically solved asexual model.}
    \label{fig:FermiPenna}
\end{center}
\end{figure}

\begin{figure}[htb]
  \begin{center}
    \includegraphics[width=0.6\textwidth]{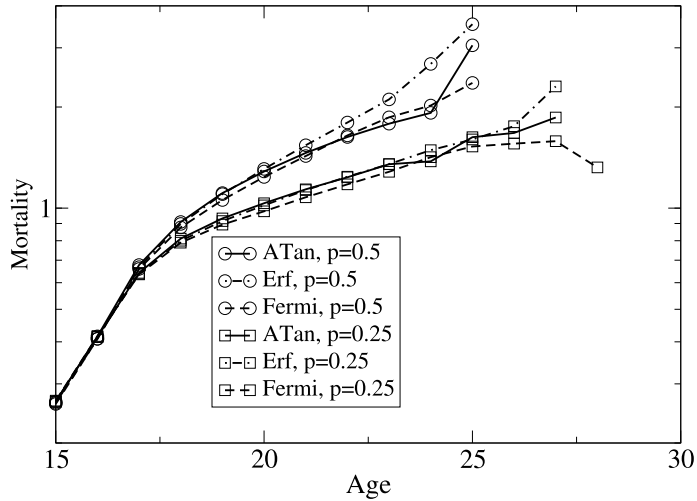}
\caption{Comparing the tails of the mortality for different 
death probability functions. They differ only slightly for different functions. 
Individuals with $n<T$ do not suffer genetic death. }
    \label{fig:CompFPenna}
\end{center}
\end{figure}

In order to obtain the mortality plateau without restricting the minimum reproduction 
age $R$, we set all values of the death probability $f(n)$ to zero for $n<T$.  
In this way the birth rate $b=1$ does not need to be increased, and the mortality for 
different values of $p$ is 
shown in Figure~\ref{fig:FermiPenna}, where the death probability is the one of 
Equation~(\ref{eq:fermi}). For young ages the mortality function 
follows the Gompertz law.
Now a nice plateau can be observed, similar to the results in~\cite{Coe2002}. 
Its length depends on the smoothness $p$. 
The different death probabilities of Equations~(\ref{eq:atan}) and (\ref{eq:erf}), also
setting to zero the genetic deaths for $n<T$, yield similar mortality functions, as shown in  
Figure~\ref{fig:CompFPenna}.

\section{Model with constant population} 

In order to study the population age structure using the death probabilities 
of Equations~(\ref{eq:fermi}) to (\ref{eq:erf}), but without neglecting deaths for $n<T$,  
we have implemented the sexual version of a model with constant population, introduced  
in \cite{Oliveira2004}. This model has the advantages of avoiding the Verhulst factor already 
criticized by many biologists and preventing chaotic fluctuations of the population size. 
The only difference between this model and the Penna one is that whenever an individual 
dies for genetic reasons, a male and a female are randomly chosen to mate and produce an 
offspring. So, the population size does not fluctuate, since there is no Verhulst factor, 
and the measured 
data are much cleaner. Additionally, the birth rate is controlled automatically and 
population meltdown or unlimited growth are prevented. Nevertheless, the simulation
can break down if there are no individuals older than the minimum reproduction age, 
which occurs for $p<1$ as well as for too small populations.
The population size (200,000 individuals) and simulation time (1,000,000 
time steps) have to be large, to produce a mortality function which ranges up
to old ages. The genetic deaths and the age distribution are measured over the last 
500,000 time steps. 

\begin{figure}[htb]
  \begin{center}
    \includegraphics[width=0.6\textwidth]{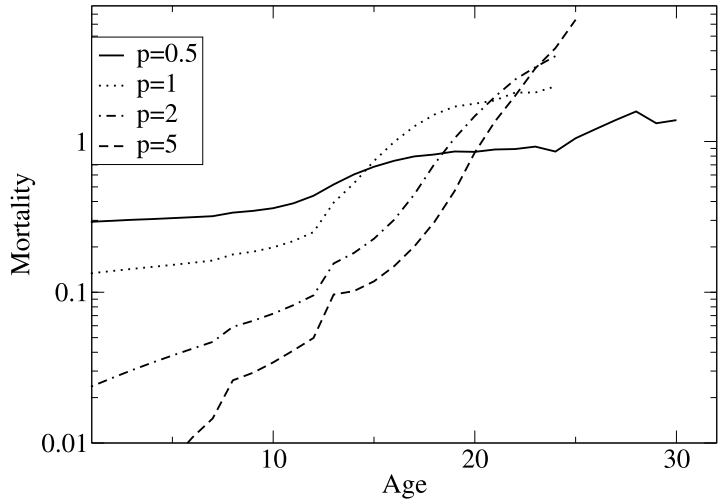}
\caption{Mortality functions using the constant population model  
with the Fermi death probability function, for different values of $p$. At young ages 
there is no Gompertz law for 
small $p$, due to the non-negligible genetic deaths for $n<T$.}
    \label{fig:Fermi}
\end{center}
\end{figure}

\begin{figure}[htb]
  \begin{center}
    \includegraphics[width=0.6\textwidth]{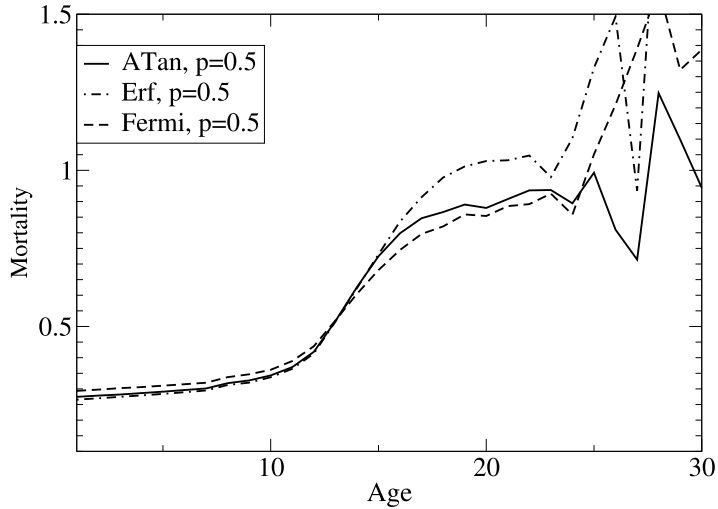}
\caption{Comparison between the mortality functions of different smooth death probabilities   
using the constant population model, in linear scale. 
The plateau appears for all of them. The fluctuations of the mortality functions for
ages above 24 result from a weak statistics.}
    \label{fig:CompF}
\end{center}
\end{figure}

Figure~\ref{fig:Fermi} shows 
that the mortality functions do not differ very much from the ones measured with 
the modified Penna model of Section 2, Figure~\ref{fig:FermiPenna}, for old ages.
But with decreasing values of $p$ the mortality increases 
considerably at young ages. The exponential growth is replaced by 
an almost constant behavior until the minimum reproduction age.
The mortality functions do not vary qualitatively for the different 
approximations of the step function (Figure~\ref{fig:CompF}), as already observed in the 
simulations of the Penna model. 

\begin{figure}[htb]
  \begin{center}
    \includegraphics[width=0.6\textwidth]{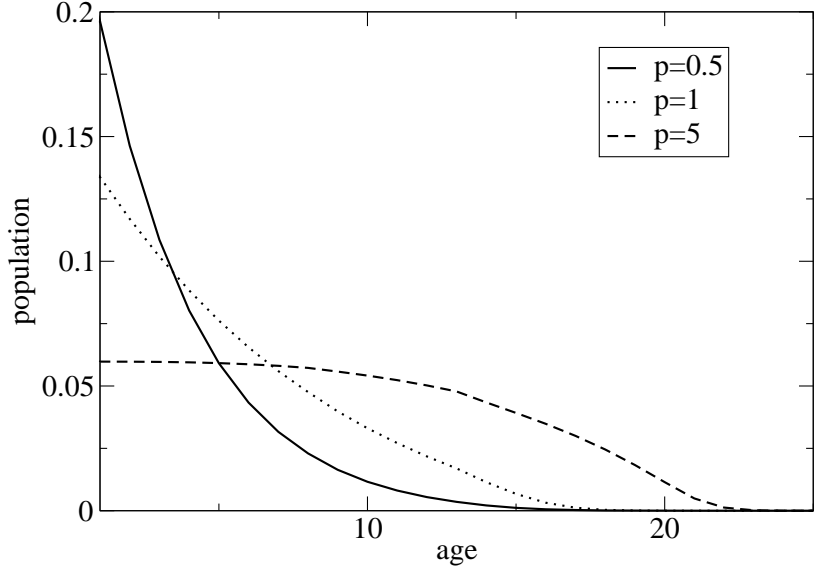}
\caption{The population density changes its curvature for small $p$. Only few individuals
reach old ages.}
    \label{fig:Pop}
\end{center}
\end{figure}

Interestingly, we observe a change in the curvature of the population age distribution, 
depending on the value of $p$ - Figure~\ref{fig:Pop}.
The smoother the death probability is, the smaller is the mean age of the population.
Most of the 
individuals die at young ages before reaching the age of reproduction. The birth rate 
increases crucially in order to maintain the population constant. The very few 
individuals who reach advanced ages can live very long. The really small number of 
these individuals explains why the mortality plateau is not observed for small 
populations or short simulation times. 
Thus, the fluctuations of the values of the mortality function at very 
old ages, shown in Figure~\ref{fig:CompF}, are due to poor statistics.

\section{Conclusion}

With our sexual simulations we reproduce the asexual results of~\cite{Coe2002}, by 
implementing a Fermi 
death probability function in the Penna model. The main differences between this  
model and the asexual model of~\cite{Coe2002} are that there, reproduction begins at birth, 
i.e. $R=1$ and the Fermi death probability function depends on the age, while in our case 
$R=10$ and the death probability depends on the current number of deleterious mutations.  

Our results 
reveal that the observation of a mortality plateau, using the traditional Penna model 
with a Fermi or any other death probability function smoother than a step function   
is a rather complicated task. For a  
reproduction age $R>1$ most of the individuals die before reaching the minimum reproduction 
age $R$. The only way to avoid population meltdown is to increase the birth rate. Simulations 
with large population size and simulation time show a small plateau. Nevertheless, 
the high chaotic fluctuations of the population size due to the large birth rate makes the 
simulations difficult. 
However, neglecting deaths before the accumulation of $T$ deleterious mutations, the model  
reproduces the Gompertz law up to old ages where the mortality function shows a 
plateau. Additionally, the birth rate does not need to be increased.  

In order to avoid neglecting deaths before the accumulation of $T$ 
mutations, we have used a constant population model. Large populations and simulation 
times also lead to a clear plateau in the mortality function, which may not follow 
the Gompertz law, depending on the value of $p$. For small $p$, many individuals die before 
reaching the reproduction age, which may change completely the population age structure.

The different approximations of the step function that have been tested in both the modified 
Penna model and the constant population model,   
have lead to similar results for old ages. Thus, we conclude that the effect of the oldest 
old results from the smoothness of the genetic death probability at old ages, within the 
theory of mutation accumulation.

The existence of plateaus in the mortality curves of Drosophilae   
and other organisms is a matter of fact, as have been reported for instance in 
\cite{Vaupel97,Vaupel98}. However, the number of Drosophilae surviving up to ages 
where the plateau appears is extremely small. This same effect has been observed with 
simulations using the constant population model, but not with the modified Penna model  
where a reasonable number of individuals survive until advanced ages. The reason 
is that to obtain the plateau with the Penna model, it is necessary to neglect deaths before 
the accumulation of $T$ mutations, which allows many individuals to survive up to the 
minimum reproduction age. The very small number of individuals reaching an age to 
observe a plateau explains the difficulty to measure the oldest old effect in Nature. 
Only experiments with more than a million of Drosophilae yield clear  
mortality plateaus, and even so, their statistics still remain quite poor.
In this way, the constant population model seems to be more 
realistic than the modified Penna model. On the other hand with the latter  
it is possible to obtain the mortality plateau without 
a great computational effort. 

Comparing the very small mortality plateau of humans with the large ones of Drosophilae, 
medflies, wasps and Nematodes~\cite{Vaupel98} we propose that there is a relation 
between the presence of a large mortality plateau and high birth rates, smooth 
death probabilities and the curvature of the population age distribution. Organisms with 
a high 
death probability at young ages need a high birth rate in order to have sufficient individuals 
reaching the reproductive age. This leads to a mortality plateau and a population 
distribution with a positive curvature. We suppose that this relation is 
valuable for simple organisms.
A similar relation between the mortality plateau and the population age distribution 
has already been observed in~\cite{Gotthard2000} for butterflies, as well as  
in~\cite{Gerhard2002} for zebrafish. 
Unfortunately, more data concerning higher developed animals are still missing. 

\bibliographystyle{revtex}
\bibliography{penna}
\end{document}